\documentclass[runningheads]{llncs}
\usepackage[T1]{fontenc}
\usepackage{graphicx}
\usepackage{amsmath}
\usepackage{tabularx}
\usepackage{booktabs} 
\usepackage{orcidlink}
\usepackage{fontawesome5}
\usepackage{comment}
\usepackage{xcolor}
\usepackage{subcaption}
\usepackage{algorithm}
\usepackage{algpseudocode}
\usepackage{makecell}

\begin{document}
%
\title{Synthetic Time Series for Anomaly Detection in Cloud Microservices}
%
%

\author{Mohamed Allam \and Noureddine Boujnah \and Noel E. O'Connor \and Mingming Liu\\}

\authorrunning{M. Allam et al.}

%

\institute{
Insight SFI Research Centre for Data Analytics, Dublin City University \\
\email{mohamed.allam@insight-centre.org}
\email{\{noureddine.boujnah,noel.oconnor,mingming.liu\}@dcu.ie}}

\maketitle              
\begin{abstract}
This paper proposes a framework for time series generation built to investigate anomaly detection in cloud microservices. In the field of cloud computing, ensuring the reliability of microservices is of paramount concern and yet a remarkably challenging task. Despite the large amount of research in this area, validation of anomaly detection algorithms in realistic environments is difficult to achieve. To address this challenge, we propose a framework to mimic the complex time series patterns representative of both normal and anomalous cloud microservices behaviors. We detail the pipeline implementation that allows deployment and management of microservices as well as the theoretical approach required to generate anomalies. Two datasets generated using the proposed framework have been made publicly available through GitHub.
\end{abstract}

\section{Introduction}
\label{sec:introduction}

The microservice architecture has emerged as a dominant paradigm for building scalable and flexible software systems. In contrast to monolithic architectures, microservices decompose applications into small, independently deployable services, each responsible for a specific business function. This architectural style offers numerous benefits, including enhanced agility, scalability, and fault isolation~\cite{ms-book}. Consequently, the adoption of microservices has witnessed a significant surge in recent years, driven by the growing demand for cloud-native and distributed systems.

Despite the advantages of microservice architectures, effectively monitoring and ensuring the reliability of microservice-based applications pose substantial challenges~\cite{nguyen2022survey}. The dynamic nature of microservices, characterized by their distributed nature and high degree of interdependence, makes traditional monitoring approaches impractical. Manual inspection becomes infeasible due to the sheer volume and complexity of services. Moreover, relying solely on anomaly detection methods focusing on single components often proves inadequate, as they fail to capture the contextual dependencies and correlations inherent in microservice environments and might lead to a very high number of false alarms~\cite{app13137891}. To address these challenges, recent research efforts have focused on developing anomaly detection techniques tailored specifically for microservice architectures. These approaches leverage multivariate analysis and incorporate diverse data modalities, combining telemetry with logs and traces, to enhance anomaly detection accuracy. 

Studies focusing on anomaly detection in microservice applications or similar distributed systems environments often utilize existing datasets~\cite{huang2023twin}~\cite{zhao2023robust} or create their own by injecting anomalies~\cite{app13137891} into simulated environments and collecting observability data. However, the scarcity of suitable datasets, compounded by the diverse array of potential anomalies and deployment configurations, can render existing datasets inadequate. Crafting an original dataset through simulating a microservice environment and intentionally introducing anomalies presents a significant challenge. This complexity arises from the need to deploy the environment, simulate a realistic load, and inject anomalies effectively. Each step requires different frameworks to interact seamlessly, demanding a high degree of integration. Additionally, while it is possible to simulate various types of anomalies, these are primarily based on assumptions and may not accurately reflect real-world production environments. Obtaining production data from cloud providers is further complicated by GDPR and other constraints. To address these challenges, we have engaged with industry experts from a world-leading cloud provider, using their insights to design the generation of load patterns and anomalies. 

In light of these challenges, we propose an anomaly generation platform tailored specifically for microservice applications leveraging Amazon Web Services (AWS)~\cite{aws} in addition to some open-source tools. Our platform encompasses a comprehensive pipeline that includes detailed descriptions of microservice deployment, load simulation, observability instrumentation, and data collection mechanisms. By providing a holistic framework for generating realistic anomaly scenarios, our platform aims to address the limitations of existing methodologies and facilitate more robust assessments of anomaly detection techniques in microservice environments through the generation of labelled multivariate datasets. In this paper, we present the design and implementation of our open-source microservice platform for anomaly generation, monitoring, and data collection. We believe that our platform will serve as a valuable tool for researchers and practitioners in the field of microservice application monitoring and anomaly detection, enabling more accurate and reliable evaluations of anomaly detection techniques in real-world settings. Furthermore, we make available two open-source labelled multivariate datasets~\footnote{https://github.com/Mohamed164/AD-microservice-app} containing some targeted anomaly scenarios. Finally, we note that the main focus of this paper is on the task of synthetic dataset generation. Due to space limitations, the application of various anomaly detection techniques on the generated datasets will be addressed in our future work.

The remainder of the paper is as follows: Section~\ref{sec:related_work} provides an overview of the current solutions in this field and highlights the key areas where further improvements are needed. Section~\ref{sec:proposed_approach} introduces our approach to address the identified gaps. In Section~\ref{sec:example_dataset}, we offer an illustrative example demonstrating the generation of a multivariate labelled dataset using our proposed approach. Section~\ref{sec:conclusions} summarizes our findings, discusses the limitations of this work, and suggests directions for future research.

\section{Related Work}
\label{sec:related_work}

In the domain of microservice anomaly detection, researchers rely on two main approaches to train data-driven machine learning algorithms and evaluate anomaly detection techniques: the creation of datasets via simulation frameworks and the utilization of pre-existing datasets. In this section, we describe these two approaches and highlight some of their advantages and shortcomings.

\subsection{Simulation Frameworks}

Microservice application simulation entails the creation of synthetic environments that closely replicate the operational characteristics of real-world microservice architectures. These simulations encompass the generation of artificial workload patterns, service interactions, and anomaly scenarios, in addition to data collection, to facilitate the evaluation of anomaly detection algorithms.
A salient advantage of simulation-based methodologies lies in their ability to include controlled experimental settings, enabling researchers to systematically manipulate parameters, introduce specific anomalies, and assess algorithmic performance under diverse conditions. Notably, Nobre et al.~\cite{app13137891} proposed a simulation framework for synthesizing time-series data representative of microservice system behavior. However, since the main focus is on the anomaly detection model the approach does not document the platform nor detail the load function used to simulate user traffic which is crucial for simulating scenarios similar to those found in the real world.


\subsection{Existing Datasets}

An alternative approach involves the utilization of publicly available datasets. These datasets typically comprise real-world operational data collected from production microservice architectures or simulated environments. Additionally, some datasets are collected from other types of distributed system architectures that are not necessarily microservices but bear significant similarities with them. These datasets are often utilized in this domain due to their relevance and applicability. By leveraging such datasets, researchers can assess algorithmic performance on authentic data and corroborate findings in real-world settings.

Jun Huang et al.~\cite{huang2023twin} trained and evaluated the performance of their anomaly detection model using two publicly available datasets. The MSDS (Multi-modal Dataset for System Anomaly Detection) \footnote{https://zenodo.org/records/226060} provides a rich resource for evaluating anomaly detection algorithms in distributed systems. The MSDS consists of distributed traces, application logs, and metrics collected from a complex distributed system (Openstack) \cite{sefraoui2012openstack} used for AI-powered analytics. The dataset includes metrics data from 5 physical nodes, each containing 7 metrics such as RAM and CPU usage, as well as log files distributed across the infrastructure with a total of 23 features. Notably, MSDS also offers a JSON file containing ground-truth information for injected anomalies, including start and end times and corresponding anomaly types, facilitating accurate evaluation of anomaly detection techniques. The second dataset is the AIOps-Challenge 2 dataset~\cite{aiops_dataset}. This dataset is derived from a simulated e-commerce system operating on a microservice architecture, with 40 service instances deployed across 6 physical nodes. It encompasses metrics recorded by each service instance, including 56 metrics, with 25 utilized in this study, covering aspects such as RAM and CPU usage. Additionally, log files are recorded for each service instance, containing a collective set of 5 features, including timestamps and original logs. The dataset traces scheduling information among service instances, capturing timestamps, types, status codes, service instance names, span IDs, parent IDs, and trace IDs. The dataset includes intentionally injected anomalies at service, pod (service instance), and node levels, accompanied by start times, levels, service names, and types. The ratio of normal to abnormal data is 120:1. A significant drawback of this dataset is that the end time of injected anomalies is not provided. 

Chenyu Zhao et al.~\cite{zhao2023robust} used the GAIA dataset \footnote{\url{https://github.com/CloudWise-OpenSource/GAIA-DataSet}}. It is a multimodal dataset collected from a system consisting of 10 instances. It consists of more than 0.7 million metrics, 87 million logs, and 28 million traces collected in a two-week period. It includes real-world injected failures alongside with their ground truth (timestamp of injection). This work also utilizes another larger dataset that is not open-source. The Server Machine Dataset (SMD) \footnote{\url{https://github.com/NetManAIOps/OmniAnomaly}} proposed by~\cite{omnianomaly} and utilized by~\cite{jumpstart}, is not a microservice dataset but rather a distributed system dataset. Spanning five weeks, it comprises data collected from 28 online service systems, each distributed across various servers. These systems offer a range of services including searching, ranking, and data processing.

While publicly available datasets confer the advantage of real-world relevance, they also pose certain challenges. A prevalent issue pertains to the restricted diversity and coverage of anomaly types within the datasets, which may not adequately encapsulate the spectrum of anomalies encountered in practice. Additionally, challenges such as data labeling inaccuracies or absent ground truth annotations can pose impediments to algorithmic evaluation and comparison.

\section{Proposed Approach}
\label{sec:proposed_approach}

In addressing the gaps highlighted in Section~\ref{sec:related_work}, our framework offers a practical approach. Firstly, it involves simulating a realistic load function that accurately mirrors the complex seasonal and trend patterns observed in real-world loads. Secondly, it enables the injection of various anomalies, each with different parameters related to user traffic load, the microservice cluster, and the underlying infrastructure. Thirdly, we propose a method to automate the labelling of these injected anomalies. Finally, we describe how metrics are collected to create a multivariate dataset, and how additional observability data, such as logs, can enhance this dataset to support multimodal models. This section details the architecture and setup of our proposed framework.

\subsection{Architecture and Main Components}

\begin{figure}[t]
  \centering
  \includegraphics[width=.8\textwidth]{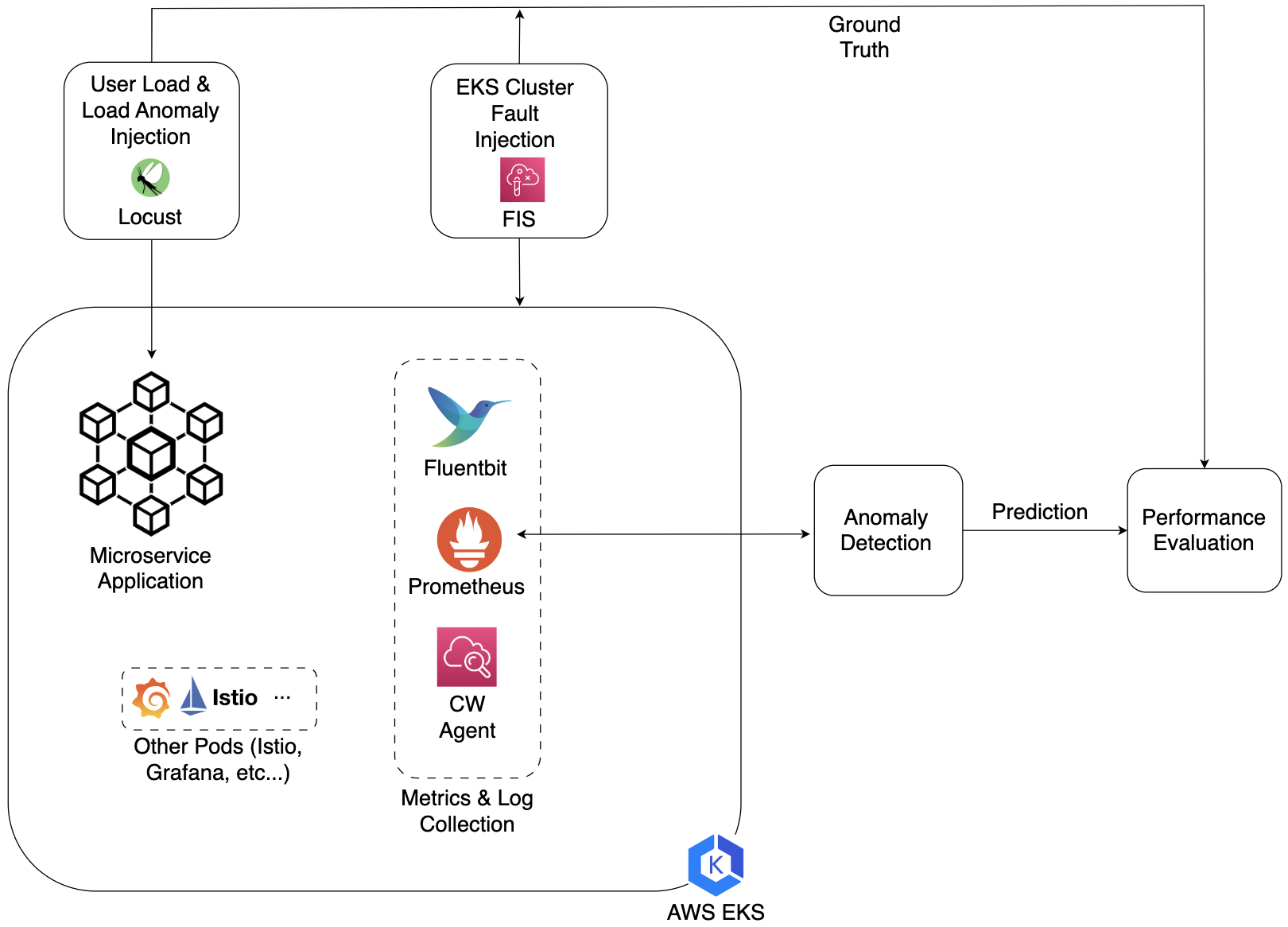}
  \caption{Experimental Setup Architecture}
  \label{fig:mvp_architecture}
\end{figure}

Our setup, depicted in Figure~\ref{fig:mvp_architecture}, involves deploying a microservice application on Amazon Elastic Kubernetes Service (EKS)~\cite{eks}, a managed Kubernetes service provided by Amazon Web Services (AWS)~\cite{aws} that simplifies deployment, management, and scaling of containerized applications. The EKS cluster is deployed on several AWS EC2 instances (Elastic Compute Cloud), which are virtual servers within the AWS cloud serving as cluster nodes. We employ two open-source applications, widely used in literature: Sock-Shop~\cite{sockshop} and Online Boutique~\cite{online-boutique}, both of which mimic e-commerce platforms with multiple microservices written in different languages and include databases. 

To inject anomalies simulating cluster malfunctions, we use AWS Fault Injection Service (FIS)~\cite{aws-fis}, for ease of integration with other AWS services deployed. For observability, we use Istio~\cite{istio}, a service mesh that manages traffic flows and collects metrics within the EKS cluster. AWS CloudWatch, a monitoring and observability service within the AWS ecosystem, is used to gather metrics from cluster nodes and pods. Additionally, the CloudWatch Fluent Bit agent is used to collect microservice logs. Prometheus~\cite{prometheus}, an open-source monitoring and alerting tool, is integrated to scrape Istio metrics, providing additional insights. 

To simulate load, we utilize Locust~\cite{locust}, a flexible load testing tool, deployed on a test server due to its ability to handle custom load functions. This setup allows us to introduce anomalies related to the application traffic.

\subsection{Load Simulation and Load Anomaly Injection}
\label{subsec:load_simulation_and_injection}

To simulate realistic user traffic patterns, we utilize the Locust framework, employing a custom function that dynamically generates users over time based on the desired number of users and their creation or spawn rate. We then explore how manipulating the user load and spawn rate functions enables the introduction of load anomalies, facilitating comprehensive testing of the system under varying conditions. Finally, within this framework, users engage in predefined scenarios that encapsulate diverse behaviors and interactions with the application, adding further realism to our simulation.

\subsubsection{Normal User Load}
\label{subsubsec:base_user_load}

To simulate the number of users, we propose an additive model incorporating multiple seasonal components, each with its own noise term. Arbitrary seasonal functions can be employed, but we suggest utilizing a sine square function. As per our conversations with experts of a leading cloud provider, this function closely mimics access patterns to services observed in some real-world datasets. The model is represented as Equation~\eqref{eq:num_users}.

\begin{equation}
\label{eq:num_users}
\text{N(t)} = (1+trend(t)) \times \text{base\_load} + \sum_{i=1}^{n} A_i \times ((\sin(\frac{2\pi t}{T_i}))^2 + wn_i(t))
\end{equation}

where:
\begin{itemize}
  \item \(A_i\) is the amplitude of the \(i^{th}\) sine squared function,
  \item \(T_i\) is the periodicity of the \(i^{th}\) sine squared function, and
  \item \(wn_i(t)\) is the noise component sampled from white noise with variance \( \sigma_i \).
\end{itemize}

and \(trend(t)\) is defined by~\eqref{eq:trend}.

\begin{equation}
\label{eq:trend}
trend(t) = \sum_{i=1}^{m} f_i(t)
\end{equation}

where $m$ is the total number of intervals, and \( f_i(t) \) is the piecewise linear function for the \( i^{th} \) interval, as defined by~\eqref{eq:f_i}.

\begin{equation}
\label{eq:f_i}
f_i(t) = 
\begin{cases} 
    \text{slope}_i, & \text{if sudden shift occurs} \\
    \text{slope}_{i-1} \cdot (1 - \frac{t - \text{start}_i}{\text{end}_i - \text{start}_i}) + \text{slope}_i \cdot \frac{t - \text{start}_i}{\text{end}_i - \text{start}_i}, & \text{otherwise} 
\end{cases}
\end{equation}

where:
\begin{itemize}
    \item \( \text{slope}_i \) is the slope of the trend within the \( i^{th} \) interval,
    \item \( \text{start}_i \) is the start of the \( i^{th} \) interval, and
    \item \( \text{end}_i \) is the end of the \( i^{th} \) interval.
\end{itemize}

The occurrence of a sudden shift within each interval \(i\) is modeled by a Bernoulli distribution with parameter \(p = 0.01\) (probability of a sudden shift in each interval), i.e., \(Y \sim \text{Bernoulli}(n, 0.01)\).

The spawn rate, \(R(t)\), is defined as the gradient of \(N(t)\) with respect to time, indicating how quickly new users are introduced into or subtracted from the system, as shown by Equation~\eqref{eq:spawn_rate}. When the spawn rate is negative, it indicates that the number of users is decreasing as virtual users are being removed.
As we are dealing with continuous and non-differentiable discrete functions, the gradient with respect to $t$ is the mean average of the left and right sides of the derivative, and provided by Equation~\eqref{eq:cont_derivative}.
 \begin{equation}
 \label{eq:cont_derivative}
     R(t)=\frac{N'_{r}(t)+N'_{l}(t)}{2}
 \end{equation}
 Where, $N'_{r}(t)$ is the right hand derivative of $N(t)$ and $N'_{l}(t)$ is the left hand derivative of $N(t)$, for the sake of computation the spawn rate can be written as Equation~\eqref{eq:spawn_rate}.
\begin{equation}
R(t)=\lim_{h \rightarrow 0}\frac{N(t+h)-N(t-h)}{2h}
\label{eq:spawn_rate}
\end{equation}

\subsubsection{Load Anomaly Generation}
\label{subsubsec:anomaly_generation}

Load anomalies are introduced into the load function by augmenting its values probabilistically, as depicted by Algorithm~\ref{alg:load_anomalies}. With a given probability \( p \), the algorithm generates a load anomaly during an interval characterized by a duration sampled from a uniform distribution \( U(a, b) \), where \( a \) and \( b \) represent the lower and upper bounds of the duration range, respectively. Additionally, the magnitude of the anomaly is determined by a multiplier sampled from a uniform distribution \( U(c, d) \), where \( c \) and \( d \) denote the lower and upper bounds of the multiplier range, respectively. This process ensures that anomalies are introduced into the load function in a controlled manner, enabling comprehensive analysis and adaptive responses to fluctuations in user activity.

\begin{algorithm}
\caption{Load Anomaly Generation}
\label{alg:load_anomalies}
\begin{algorithmic}[1]
\State $anomaly\_end\_time \gets$ None
\State $anomaly\_multiplier \gets$ None
\For{$(timestamp, value)$ in $user\_load$}
\If{$anomaly\_end\_time$ \textbf{and} $timestamp \leq anomaly\_end\_time$}
\State Mark $timestamp$ as anomaly
\State Adjust $value$ using $anomaly\_multiplier$
\ElsIf{random number $< p$}
\State $anomaly\_duration \gets U(a, b)$ \Comment{Sample anomaly duration}
\State $anomaly\_end\_time \gets timestamp + anomaly\_duration$
\State Mark $timestamp$ as anomaly
\State $anomaly\_multiplier \gets U(c, d)$ \Comment{Sample anomaly multiplier}
\State Adjust $value$ using $anomaly\_multiplier$
\EndIf
\EndFor
\end{algorithmic}
\end{algorithm}

\begin{figure}[t]
    \centering
    \begin{subfigure}[b]{\textwidth}
        \centering
        \includegraphics[width=0.8\textwidth]{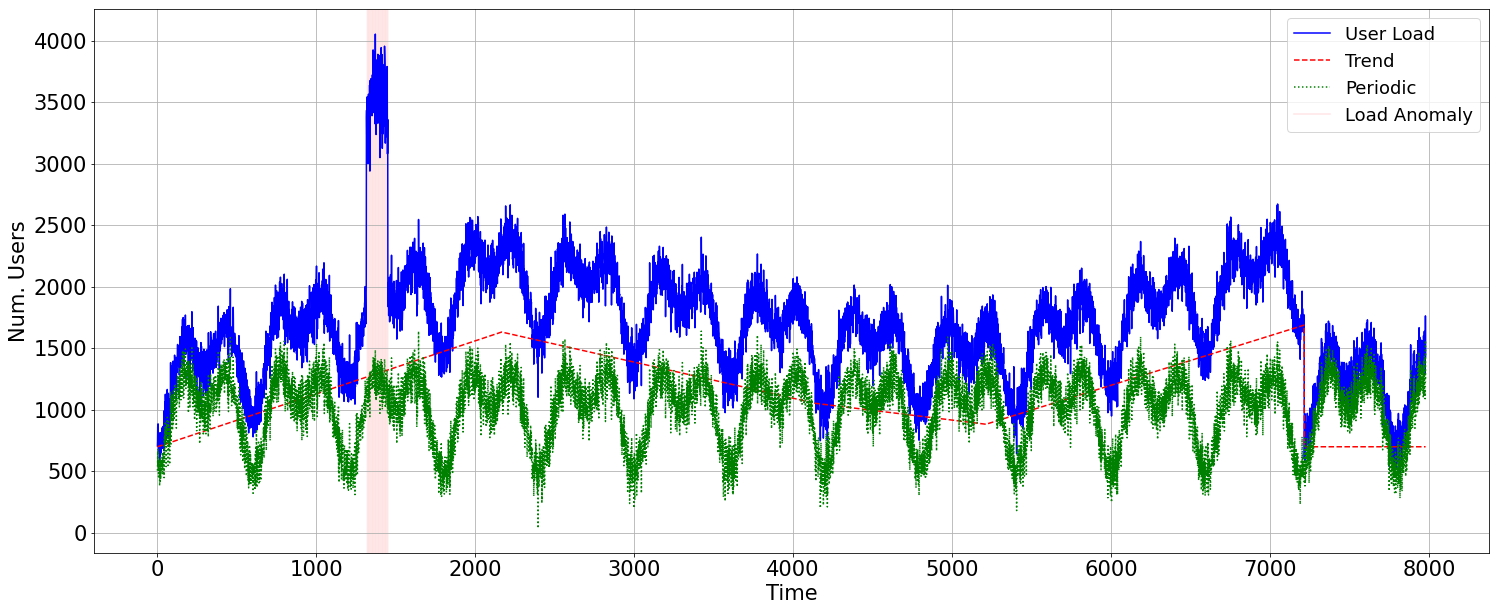}
        \caption{Load Function}
        \label{fig:load_func}
    \end{subfigure}
    
    \begin{subfigure}[b]{\textwidth}
        \centering
        \includegraphics[width=0.8\textwidth]{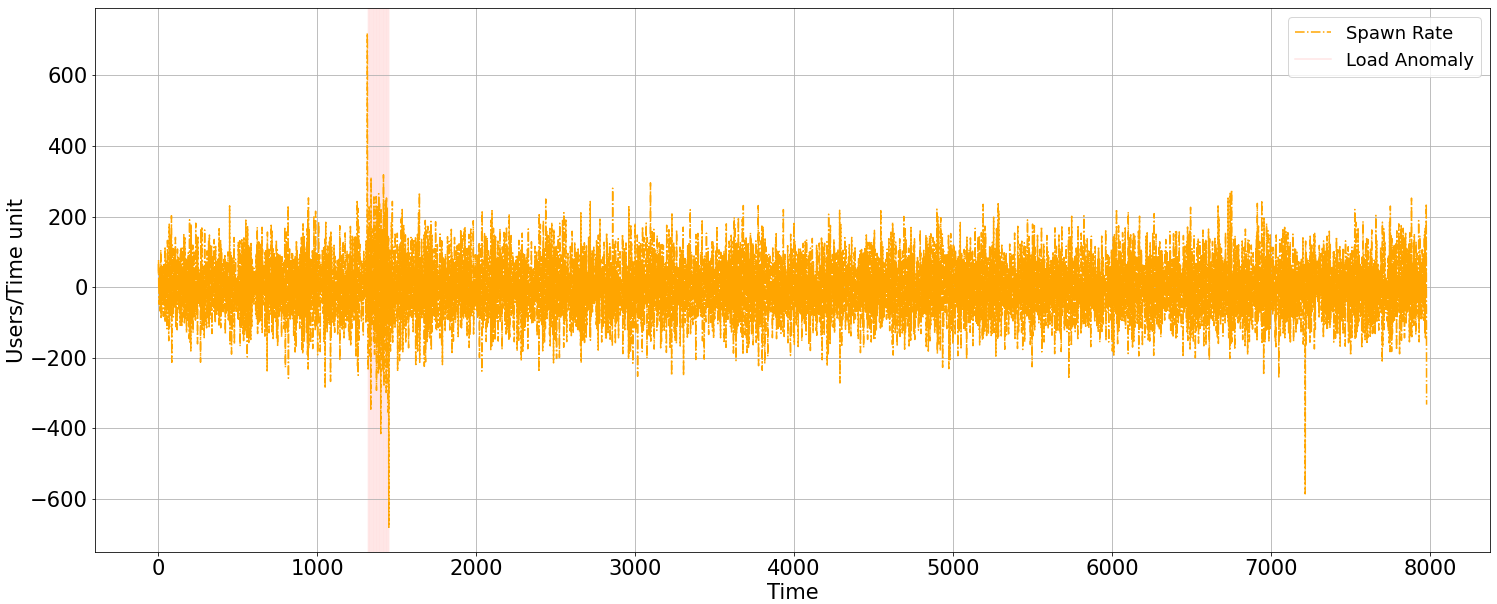}
        \caption{Spawn Rate}
        \label{fig:spawn_rate}
    \end{subfigure}
    \caption{Load Function and Spawn Rate}
    \label{fig:load_and_spawn}
\end{figure}

Figure~\ref{fig:load_and_spawn} illustrates an example of a load function generated using the proposed framework. The trend and periodic components are decomposed. It includes increasing and decreasing trend intervals in addition to a sudden shift. The spawn rate is also depicted in the figure. 

\subsubsection{User Scenarios}
\label{subsubsec:user_scenarios}

To effectively simulate user load on both applications, which are e-commerce websites with similar functionalities, we define three distinct user scenarios:
\begin{itemize}
    \item Visitor User: a user that visits the homepage and catalogue
    \item  New Shopper User: user that registers, visits catalogue, adds a random item to the cart, and makes the order.
    \item Returning Shopper User: a similar flow to the new shopper users but the user is already registered so performs only a login.
\end{itemize}

\begin{figure}[t]
  \centering
  \includegraphics[width=1.0\textwidth]{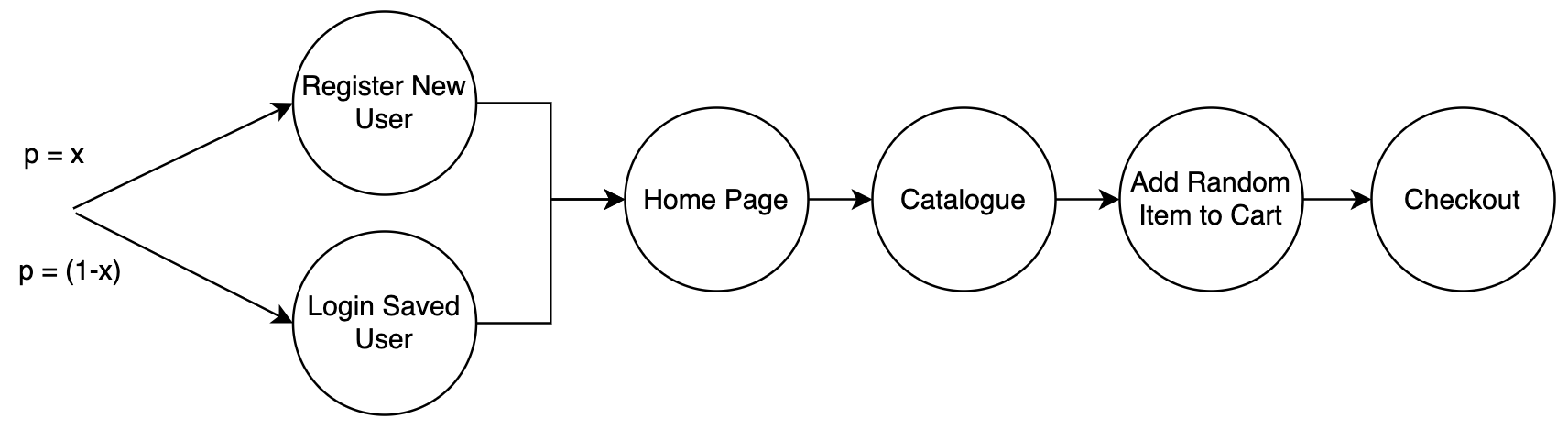}
  \caption{User Scenarios: New Shopper User and Returning Shopper User}
  \label{fig:user_scenarios}
\end{figure}

Figure~\ref{fig:user_scenarios} illustrates a scenario that combines the actions of New Shopper User and Returning Shopper User. The probability of a New Shopper User workflow is denoted by 
$x$, while the probability of a Returning Shopper User workflow is 1-$x$.

\subsection{EKS Cluster Anomaly Injection}

FIS is employed to systematically introduce anomalies within our EKS cluster. These scenarios are defined as templates and encompass various anomaly types at both the node and pod levels.
At the pod level, anomalies include pod deletion, CPU stress, memory stress, increased network latency, packet dropping, and I/O stress. Similarly, at the node level, anomalies involve CPU stress, memory stress, increased network latency, packet dropping, I/O stress, and instance rebooting. Table~\ref{tab:fis_anomalies} summarizes these anomalies and their parameters.


\begin{table}[t]
\caption{Parameters of FIS experiments}
\label{tab:fis_anomalies}
\centering
\begin{tabularx}{\textwidth}{l|l|l}
\toprule
\textbf{Anomaly} & \textbf{Target} & \textbf{Parameters}  \\
\midrule
Pod Deletion & pod & target\\
\midrule
CPU Stress & pod/node & target, duration, load \% \\
\midrule
Memory Stress & pod/node & target, duration, load \% \\
\midrule
Network Latency & pod/node & target, duration, latency magnitude \\
\midrule
Packet Dropping & pod/node & target, duration, packet drop rate  \\
\midrule
I/O Stress & pod/node & target, duration, I/O space \% \\
\midrule
Node Reboot & node & target \\
\bottomrule
\end{tabularx}
\end{table}

Anomalies are deployed by defining several anomaly templates on AWS FIS. Each anomaly template requires the definition of an action, for example, CPU stress, with its parameters, and a target, for example, an EC2 instance or a pod. We define a total of 12 templates for every anomaly and target type in Table~\ref{tab:fis_anomalies}. The targets are defined as "ec2-target" or "pod-target" depending on the target type and are left as placeholders. In the case of EC2 anomalies, we use the Amazon Resource Name (ARN) of the target instance, and in the case of pods, we use a label selector to identify the pods to be targeted. We use an IAM role with policies allowing access to the EKS cluster, EC2 instances and CW for logging. We also setup log forwarding to CW as these logs will serve as ground truth for anomaly labelling in a later stage.
We then use a Python script, using the Boto3 library~\cite{boto3}, to deploy these experiments according to a probability using a uniform distribution. Deployment consists of two steps: updating the FIS experiment template by substituting the placeholder with a random target or multiple targets, and then deploying the experiment. Algorithms~\ref{alg:anomaly_deployment} summarizes these steps.

\begin{algorithm}
\caption{Deploying Anomaly Experiments}
\label{alg:anomaly_deployment}
\begin{algorithmic}[1]
\Require Probability of anomaly $p_{\text{anomaly}}$, List of anomaly templates
\While{True}
    \If{$\text{random}() < p_{\text{anomaly}}$}
        \State Pick a random anomaly template
        \If{Target type is EC2}
            \State Use ARN of random target instance(s)
        \Else
            \If{Target type is pod}
                \State Use label selector to target random pod(s)
            \EndIf
        \EndIf
        \State Update experiment template
        \State Deploy experiment
    \EndIf
    \State Wait for next minute
\EndWhile
\end{algorithmic}
\end{algorithm}

\subsection{Metrics Collection}

\begin{figure}[t]
  \centering
  \includegraphics[width=1.0\textwidth]{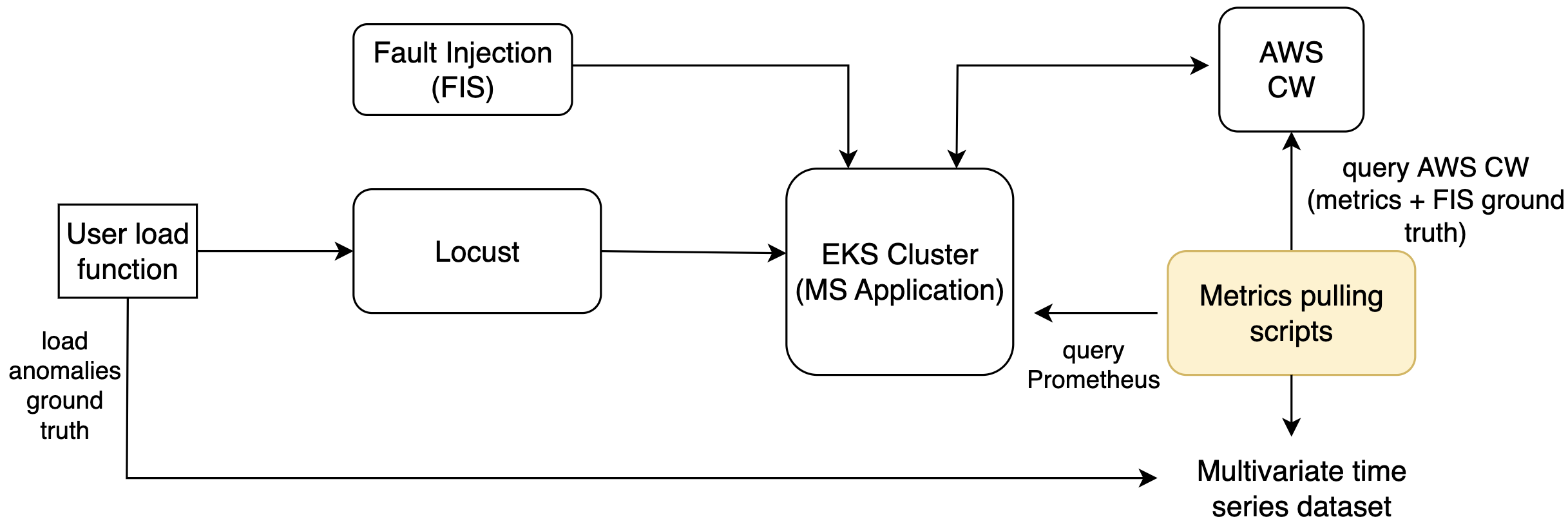}
  \caption{Metrics Collection}
  \label{fig:metrics_collection}
\end{figure}

To collect metrics we utilize the Boto3 library and the Prometheus API, in the highlighted step in Figure~\ref{fig:metrics_collection}. Node and pod-level metrics are obtained by querying CloudWatch, while service-level metrics (coming from Istio) are retrieved from Prometheus. Load anomaly ground truth is logged by the Locust custom script, while FIS experiments' information and ground truth are retrieved from its CloudWatch logs. By combining these datasets, we obtain a comprehensive, multivariate dataset describing the state of the cluster, its nodes, pods, and services. This dataset includes anomaly injection ground truths, making it a valuable resource for training and evaluating anomaly detection algorithms.

The final dataset comprises node-level metrics, pod-level metrics and service-level metrics. In addition to the anomaly ground truth composed of FIS experiment information (experiment deployed, start and end time), and load anomaly injection information (start time, end time). 


\subsection{Logs Collection}

Logs are collected by the CloudWatch agent\footnote{https://docs.aws.amazon.com/eks/latest/userguide/eks-observe.html} and forwarded to CloudWatch where they are saved. These logs can be queried by utilizing the Boto3 library. There are four types of log groups generated by the CloudWatch agent:

\begin{itemize}
    \item \textbf{/aws/containerinsights/<cluster\_name>/dataplane}: This log group contains logs related to the data plane of your Amazon EKS cluster. It includes information about network traffic, load balancer activity, and communication between nodes.
    \item \textbf{/aws/containerinsights/<cluster\_name>/application}: Logs relevant to the applications running on the cluster are stored in this log group. It includes application-level logs, such as logs generated by services or applications deployed within the cluster.
    \item \textbf{/aws/containerinsights/<cluster\_name>/performance}: This log group captures performance-related logs, providing insights into the resource utilization, latency, and other performance aspects of the cluster.
    \item \textbf{/aws/containerinsights/<cluster\_name>/host}: Logs related to the underlying hosts or nodes of the cluster are stored in this log group. It includes system-level logs, such as kernel messages, hardware events, and other host-specific information.
\end{itemize}


\section{Example Dataset}
\label{sec:example_dataset}

This section presents a dataset created using our proposed framework. The parameters below were used as inputs in our dataset generation pipeline for both published datasets.

The following parameters were used as input to the Locust load function, \( N(t) \), as defined by Equation~\ref{eq:num_users}: the periodic functions \( (600, 50, 0.1) \), \( (320, 40, 0.05) \), \( (30, 9, 0.01) \), and \( (60, 16, 0.05) \); a total duration of 7 days (\( \text{duration} = 604800 \)); the number of intervals for trend computation (\( \text{num\_intervals} = 30 \)); the range of mean slopes (\( \text{slope\_mean\_range} = (-0.98, 2.5) \)); the probability of sudden shifts (\( \text{prob\_sudden\_shift} = 0.0001 \)); the baseline load (\( \text{base\_load} = 20.0 \)); the anomaly probability (\( \text{anomaly\_prob} = 0.00025 \)); and the range of anomaly multipliers (\( \text{anomaly\_multiplier\_range} = (1, 2) \)). Table~\ref{tab:experiment_parameters} provides a summary of the experiment parameters used to inject EKS cluster anomalies.

\begin{table}[htbp]
\centering
\caption{FIS Experiment Parameters}
\label{tab:experiment_parameters}
\begin{tabular}{lll}
\toprule
\textbf{Anomaly} & \textbf{Parameters}  \\
\midrule
CPU Stress & \makecell[l]{Load perc.: 100\%, Duration: 4 mins} \\
Memory Stress & \makecell[l]{Load perc.: 100\%, Duration: 4 mins}  \\
Network Latency & \makecell[l]{Delay: 400 ms, Duration: 5 mins}  \\
Packet Dropping & \makecell[l]{Loss Perc.: 40\%, Duration: 2 mins} \\
I/O Stress & \makecell[l]{Load perc.: 80\%, Duration: 5 mins} \\
Pod Deletion & - \\
Node Reboot & - \\
\bottomrule
\end{tabular}
\end{table}

The structure and dimensionality of the metrics dataset are summarized in Table~\ref{tab:dataset_structure}. This dataset includes all collected metrics along with the labeled ground truth. Additionally, we provide a separate dataset containing information about injected FIS anomalies, which includes their types, targets, and start and end timestamps.

\begin{table}[htbp]
\centering
\caption{Dataset Structure Summary (G.T.: ground truth)}
\label{tab:dataset_structure}
\begin{tabular}{lllr} 
\toprule
\textbf{Entity Type} & \textbf{Metrics} & \textbf{\#Entities} & \textbf{Total}\\ 
\midrule
Node & 25 & 5 & 125 \\ 
Pod & 18 & 14 & 252\\ 
Service & 9 & 10 & 90 \\ 
Load anomaly G.T. & 1 & - & 1 \\ 
FIS anomaly G.T. & 1 & - & 1 \\ 
Anomaly G.T. & 1 & - & 1 \\ 
\bottomrule
\textbf{Total} &  &  & \textbf{470} \\ 
\bottomrule
\end{tabular}
\end{table}

\begin{figure}[h!]
  \centering
  \includegraphics[width=0.9\textwidth]{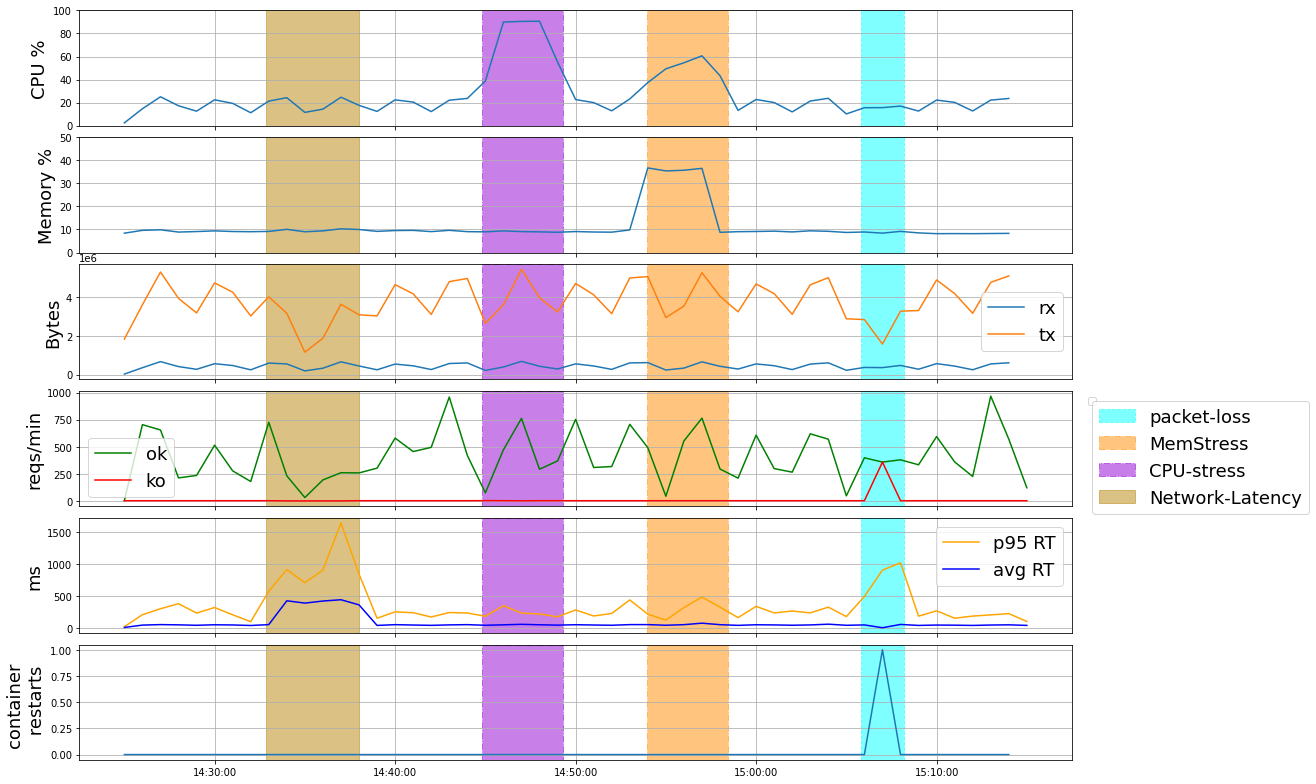}
  \caption{Effect of various anomalies deployed on the Front-end pod on key metrics}
  \label{fig:front_end_anomalies}
\end{figure}

Figure~\ref{fig:front_end_anomalies} illustrates the impact of various anomalies deployed on the Front-end on its key metrics. The network-latency anomaly prominently increases both the average and 95th percentile response times of the Front-end service metrics. Additionally, there is a slight decrease in the received bytes metric and a drop in the request rate. However, metrics such as CPU and memory utilization remain unaffected.
The CPU stress anomaly significantly elevates the CPU percentage metric. Despite the CPU percentage occasionally hitting 100\%, container or pod restarts are not observed in this scenario.
On the other hand, the memory stress anomaly results in heightened memory and CPU percentages but does not cause failures or impact user experience.
In contrast, the packet loss anomaly leads to substantial failures, manifesting in increased p95 response times and the number of failed requests. Ultimately, this anomaly necessitates a container restart.

\begin{figure}[h!]
  \centering
  \includegraphics[width=0.9\textwidth]{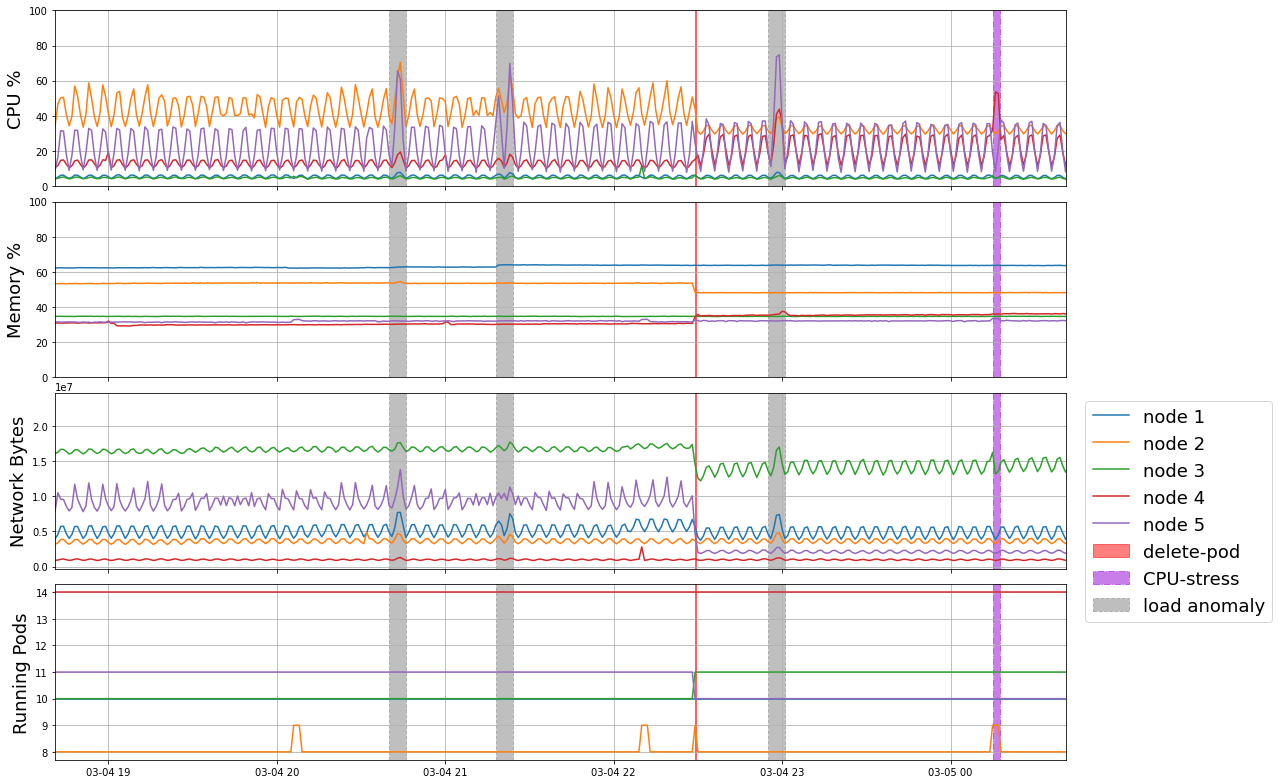}
  \caption{Effect of various anomalies on key cluster node metrics [load anomalies, CPU stress on Front-end pod, Front-end pod deletion]}
  \label{fig:ec2_anomalies}
\end{figure}

Figure~\ref{fig:ec2_anomalies} illustrates the effects of three types of anomalies on key EC2 metrics of the 5 nodes of the EKS cluster. Load anomalies cause an evident increase in the CPU metrics of various cluster nodes. The same applies to the network bytes metric, while memory metrics are relatively unaffected. The CPU stress anomaly, deployed on the Front-end pod, causes a spike in the CPU utilization metric of the node on which the pod is deployed (node 4). Pod deletion causes significant changes in all metrics, as the pod is restarted on a different node. Thus, we notice a drop in the metrics of node 2 and an increase in those of node 4. Pod deletion is a case of how an error on a certain cluster node affects metrics of a different cluster node.

\section{Conclusions and Future Work}
\label{sec:conclusions}

This paper provides an in-depth overview of a platform designed for deploying microservice applications in a distributed cluster, simulating user load, and injecting various types of anomalies. The platform's effectiveness is demonstrated by generating a multivariate labeled dataset specifically tailored for anomaly detection tasks within a distributed microservice environment. We have described in detail the pipeline implementation that facilitates the deployment and management of microservices, along with the methods required to generate anomalies, label them, and collect observability data. Additionally, we made publicly available two datasets generated using the proposed framework.

Despite the platform's strengths, there are some limitations in this work. While we propose a modular framework capable of supporting various types of microservice applications, our current deployment is limited to e-commerce applications. Additionally, while we gather metrics and logs, we do not include traces in our observability data at present. Traces, which track request propagation across services, are sometimes crucial for detailed error analysis and will be incorporated in our future work. Moreover, our discussion currently focuses only on synthetic data generation for anomaly detection. Future efforts will explore how to leverage our framework and datasets for anomaly detection tasks, including a comparative analysis of different state-of-the-art data-driven anomaly detection algorithms.

\section*{Acknowledgment}
This research was conducted with the financial support of Science Foundation Ireland \textit{12/RC/2289\_P2} at Insight the SFI Research Centre for Data Analytics at Dublin City University. The data used in this work is fully anonymous. For the purpose of Open Access, the author has applied a CC BY public copyright licence to any Author Accepted Manuscript version arising from this submission.

\newpage

%
%
\bibliographystyle{splncs04}
\bibliography{biblio}

\end{document}